# Surgical Data Science: A Consensus Perspective

Lena Maier-Hein, Matthias Eisenmann, Carolin Feldmann, Hubertus Feussner, Germain Forestier, Stamatia Giannarou, Bernard Gibaud, Gregory D. Hager, Makoto Hashizume, Darko Katic, Hannes Kenngott, Ron Kikinis, Michael Kranzfelder, Anand Malpani, Keno März, Beat Müller-Stich, Nassir Navab, Thomas Neumuth, Nicolas Padoy, Adrian Park, Carla Pugh, Nicolai Schoch, Danail Stoyanov, Russell Taylor, Martin Wagner, S. Swaroop Vedula, Pierre Jannin*, and Stefanie Speidel*

L. Maier-Hein, M. Eisenmann, C. Feldmann, and K. März are with the Division Computer Assisted Medical Interventions (CAMI), German Cancer Research Center (DKFZ), 69120, Heidelberg, Germany.

A. Malpani and S. Vedula are with the Malone Center for Engineering in Healthcare, The Johns Hopkins University, Baltimore, MD, 21218, USA.

H. Feussner and M. Kranzfelder are with the Department of Surgery, Klinikum rechts der Isar, Technical University of Munich, 81675, Munich, Germany.

G. Forestier is with IRIMAS, University of Haute-Alsace, 68093, Mulhouse, France.

S. Giannarou is with the Hamlyn Centre for Robotic Surgery, Imperial College London, London, SW7 2AZ, UK.

B. Gibaud and P. Jannin are with the University of Rennes 1, 35065, Rennes, France, and with INSERM, 35043, Rennes, France.

G. Hager is with the Malone Center for Engineering in Healthcare, The Johns Hopkins University, Baltimore, MD, 21218, USA, and with the Department of Computer Science, The Johns Hopkins University, Baltimore, MD, 21218, USA.

M. Hashizume is with the Department of Advanced Medical Initiatives, Graduate School of Medical Sciences, Kyushu University, Fukuoka, 812-8582, Japan.

D. Katic is with the Institute for Anthropomatics and Robotics, Karlsruhe Institute of Technolgoy (KIT), 76131, Karlsruhe, Germany.

R. Kikinis is with the Department of Radiology, Brigham and Women's Hospital and Harvard Medical School, Boston, MA, 02215, USA, with the Department of Computer Science, University of Bremen, 28359, Bremen, Germany, and with Fraunhofer MEVIS, 28359, Bremen, Germany.

B. Müller-Stich, H. Kenngott and M. Wagner are with the Department for General, Visceral and Transplant Surgery, Heidelberg University Hospital, 69120, Heidelberg, Germany.

N. Navab is with the Chair for Computer Aided Medical Procedures, Technical University of Munich, 80333, Munich, Germany, and with the Department of Computer Science, The Johns Hopkins University, Baltimore, MD, 21218, USA.

T. Neumuth is with the Innovation Center Computer Assisted Surgery (ICCAS), University of Leipzig, 04103, Leipzig, Germany.

N. Padoy is with ICube, University of Strasbourg, CNRS, IHU, 67081, Strasbourg, France.

A. Park is with the Department of Surgery, Anne Arundel Health System, Annapolis, MD, 21401, USA, and with the Johns Hopkins University School of Medicine, Baltimore, MD, 21205, USA.

C. Pugh is with the Department of Surgery, University of Wisconsin, Madison, WI, 53792, USA.

N. Schoch is with the Engineering Mathematics and Computing Lab (EMCL), IWR, Heidelberg University, 69120, Heidelberg, Germany.

D. Stoyanov is with the Centre for Medical Image Computing (CMIC) and Department of Computer Science, University College London, London, WC1E 6BT, UK.

R. Taylor is with the Department of Computer Science, The Johns Hopkins University, Baltimore, MD, 21218, USA.

S. Speidel is with the Division Translational Surgical Oncology, National Center for Tumor Diseases (NCT), 01307, Dresden, Germany.

*: Equal contribution senior authors

*Abstract*—Surgical data science is a scientific discipline with the objective of improving the quality of interventional healthcare and its value through capturing, organization, analysis, and modeling of data. The goal of the *1st workshop on Surgical Data Science* was to bring together researchers working on diverse topics in surgical data science in order to discuss existing challenges, potential standards and new research directions in the field. Inspired by current open space and think tank formats, it was organized in June 2016 in Heidelberg. While the first day of the workshop, which was dominated by interactive sessions, was open to the public, the second day was reserved for a board meeting on which the information gathered on the public day was processed by (1) discussing remaining open issues, (2) deriving a joint definition for surgical data science and (3) proposing potential strategies for advancing the field. This document summarizes the key findings.

## I. INTRODUCTION

DATA science is an interdisciplinary field that "is expected to make sense of the vast stores of big data" [1]. While no consensus definition has been established [2], researchers agree that it is a dedicated field whose core research objectives differ from that of established branches, such as natural or social sciences. With the recent advances in artificial intelligence, data science techniques have been affecting various domains, including machine translation, speech recognition, robotics, and search engines and heavily influence economics, business and finance today.

While various subfields, like "biological data science", "social data science", "business data science" and many others have been introduced in the past years, the application of data science to interventional medicine (e.g. surgery, interventional radiology, radiation therapy) has found almost no attention in the literature. This can partly be attributed to the fact that only a fraction of patient-related data and information is being digitized and stored in a structured manner. Furthermore, perioperative care has traditionally been based on local traditions, experience or individual preferences of physicians and other staff. However, with the increasing availability of (interventional) imaging data as well as recent developments in the fields of computer-assisted interventions and personalized medicine, data science is now evolving to be a key enabling technique to support clinicians in interventional disease diagnosis and therapy, paving the way for knowledge-based rather than "eminence-based" healthcare.

The purpose of this paper is to provide a comprehensive, consensus perspective on the emerging scientific discipline of Surgical Data Science (SDS), as well as to provide a roadmap for advancing the field. It is based on an interactive workshop[1], organized in June 2016 in Heidelberg that brought together leading international researchers working on diverse topics in SDS. Section II reviews the workshop format with a particular focus on the interactive parts that served as the basis for this paper. Section III reviews the collaborative definition of SDS developed at the workshop (as published in [3]) and provides an overview of associated technical research fields as well as of the range of clinical applications. The following two sections IV and V provide a review of the opportunities and challenges associated with SDS, highlighting related research fields as well as existing initiatives and standards. Based on the opportunities and challenges identified, section VI presents a joint strategy towards advancing the field, including a prioritization of technical challenges and clinical applications, strategies with respect to shared data repositories and ontologies as well as political considerations.

## II. METHODS
AUTHORS: L. MAIER-HEIN, M. EISENMANN, K. MÄRZ

This review is based on the first international workshop on SDS, which was funded by the Collaborative Research Center Cognition-guided Surgery (SFB/Transregio 125) and hosted by the German Cancer Research Center (DKFZ) in Heidelberg in June 2016. Inspired by current open space and think tank formats, it took place on two days.

1) **Public day**: The first part of the workshop was open to the public (free registration) and announced via various mailing lists. It was attended by about 75 participants from Asia, North America, and Europe. On this day, keynote lectures by leading experts in the field were complemented by short presentations of accepted workshop papers. The core component of the first workshop day were two interactive sessions (cf. sections II-A and II-B) that served as the foundation for the present paper.
2) **Board meeting**: On the second day of the workshop the board members of the workshop[2] met with the goal of further processing the information gathered on the public day by (1) discussing remaining open issues, (2) deriving a joint definition for SDS and (3) proposing potential strategies for advancing the field.

The following subsections present the workshop's concepts for the two interactive sessions, the purpose of which was to review the new field of SDS and to discuss new strategies for advancing the field. The results of the interactive sessions and the second day meeting are reflected in Sec. III to V.

[1] www.surgical-data-science.org/workshop2016
[2] www.surgical-data-science.org/workshop2016/committee

TABLE I
BRAINWRITING QUESTIONS TO DEFINE AND REVIEW THE FIELD OF SDS

| | |
|---|---|
| Q1 | What is SDS? |
| Q2 | What are distinguishing properties of SDS compared to general data science? |
| Q3 | What do you consider to be key clinical applications of SDS? |
| Q4 | What are technical research fields in SDS? |
| Q5 | What do you regard as key technical challenges in SDS? |
| Q6 | What do you consider to be moral, ethical or social challenges related to SDS? |
| Q7 | What do you consider to be the key initiatives/projects/standards in the field? |
| Q8 | What are related domains/research fields? |
| Q9 | Do you have questions? |
| Q10 | Which topics do you want to discuss in order to move the field of SDS forward? |

### A. Brainwriting for reviewing the field

Brainwriting [4] is a technique for gathering information, solving problems or generating ideas in a group. In contrast to brainstorming, which is typically face-to-face and serial in nature, it allows participants to bring in ideas/information in parallel and anonymously via writing. In the variant chosen for the workshop, all participants were asked to pin Post-It notes with keywords/ideas to poster walls representing dedicated questions (Tab. II-A and Fig. 1).

In a second round, the participants received green and red stickers to express strong agreement (green) and disagreement (red) with the notes posted by other participants. Finally, dedicated workshop participants were asked to group the Post-Its, as shown in Fig. 1.

### B. World Café for defining a joint roadmap

The World Café methodology [5] is a workshop method for hosting large group dialogue. While various variants exist we chose the following format:

1) **Questions**: Based on the brainwriting sessions (especially Q9 and Q10 in Tab. II-A), eight questions to be discussed by the workshop participants were selected by the workshop board, as summarized in Tab. II-B.
2) **Setting**: Eight tables, each "hosted" by one workshop board member and representing one of the eight questions, were set up, as illustrated in Fig. 1.
3) **Small group rounds**: Three twenty minute rounds of conversation were held. Before each round, each workshop participant chose one of the tables (and thus one dedicated topic) for discussion. The host introduced the table question and (in the second and third round) summarized the discussions of the previous rounds before moderating the new round.
4) **Plenum presentation**: After the three rounds, the table hosts summarized the discussions held at their tables, and further questions of the audience were discussed in a plenum format.

During the board discussions on the second (closed) day of the workshop, the input of the workshop participants was used (Fig. 1) in developing a consensus definition of SDS (Sec. III) and for generating an outline for the present paper that reflects both the joint review of the field (Sec. IV: Opportunities; Sec. V: Challenges) as well as a joint roadmap for advancing the field (Sec. VI).

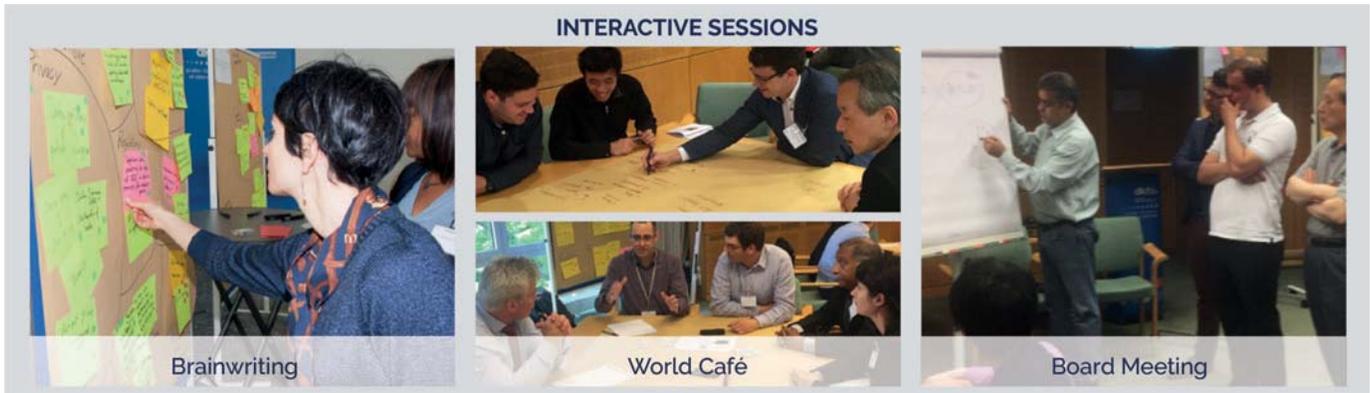

Fig. 1. Interactive sessions of the workshop.

TABLE II
WORLD CAFÉ QUESTIONS TO DEFINE JOINT STRATEGIES FOR ADVANCING THE FIELD OF SDS

| | | |
|---|---|---|
| W1 | **Key research questions**: | What are the key (highest priority) research questions? |
| W2 | **Key clinical applications**: | What are the high impact clinical applications short-term, mid-term, long-term? |
| W3 | **Shared ontologies**: | Do we need a common ontology? What do we have? How do we proceed? |
| W4 | **Shared data access**: | How can we create an international clinical data repository? |
| W5 | **Shared tools and software**: | What do we need? What do we have? |
| W6 | **Validation**: | How can we ensure validity of findings in surgical data analysis? |
| W7 | **Incentives**: | How can we convince clinicians and other stakeholders to invest in SDS? |
| W8 | **Dissemination**: | How can we translate research results into clinical practice |

### III. WHAT IS SDS?

The paradigm of SDS is illustrated in Fig. 3. Section III-A reviews the joint definition of SDS derived at the workshop and explains the technical modules involved in an SDS system (Fig. 3 "Continuous Learning"), while Sec. III-B reviews the range of clinical applications (Fig. 3 "Application").

#### A. Definition
*Authors: L. Maier-Hein, K. März, C. Feldmann and P. Jannin*

Based on brainwriting questions Q1 and Q2 (Tab. II-A) as well as extensive discussions held during the workshop day, the board members agreed that SDS deals with the *manipulation of a target anatomical structure to achieve a specified clinical objective during patient care*. In contrast to general biomedical data science, it also includes procedural data, involving the four main components depicted in Fig. 2.

It was further agreed that SDS should not only be related to surgery but also to other disciplines that deal with interventional disease diagnosis and treatment. In particular, interventional radiology and radiation therapy are included within its scope. The prefixes surgical/interventional and procedural were discussed, however, "surgical" was chosen because (1) it has a clear relation to medicine, (2) it is highly integrative, with various medical disciplines being part of the perioperative process, (3) surgeons tend to relate the term "interventional" to other disciplines. The following consensus definition of SDS was agreed on [3]:

*SDS aims to improve the quality of interventional healthcare and its value through the capture, organization, analysis and modeling of data. It encompasses all clinical disciplines in which patient care requires intervention to manipulate anatomical structures with a diagnostic, prognostic or therapeutic goal, such as surgery, interventional radiology, radiotherapy, and interventional gastroenterology. Data may pertain to any part of the patient-care process (from initial presentation to long-term outcomes), may concern the patient, caregivers, and/or technology used to deliver care, and is analyzed in the context of generic domain-specific knowledge derived from existing evidence, clinical guidelines, current practice patterns, caregiver experience and patient preferences. Data may be obtained through medical records, imaging, medical devices or sensors that may either be positioned on patients or caregivers, or integrated into the instruments and technology used to deliver care. Improvement may result from understanding processes and strategies, predicting events and clinical outcome, assisting physicians in decision-making and planning execution, optimizing ergonomics of systems, controlling devices before, during and after treatment, and from advances in prevention, training, simulation and assessment. SDS builds on principles and methods from other data-intensive disciplines, such as computer science, engineering, information theory, statistics, mathematics and epidemiology, and complements other information-enabled technologies such as surgical robotics, smart operating rooms and electronic patient records.*

In a nutshell, SDS aims to observe everything that occurs within and around the treatment process in order to provide the right assistance to the right person at the right time. The core component of an SDS system is the knowledge base (Fig. 3), which contains all the available domain knowledge. The latter can be classified into factual and practical knowledge [6]:

Factual knowledge has been written down in quotable sources. Prominent examples include clinical guidelines, studies, and educational books. In contrast, practical knowledge results from experience. It comprises case knowledge as well as expert knowledge, opinions and preferences.

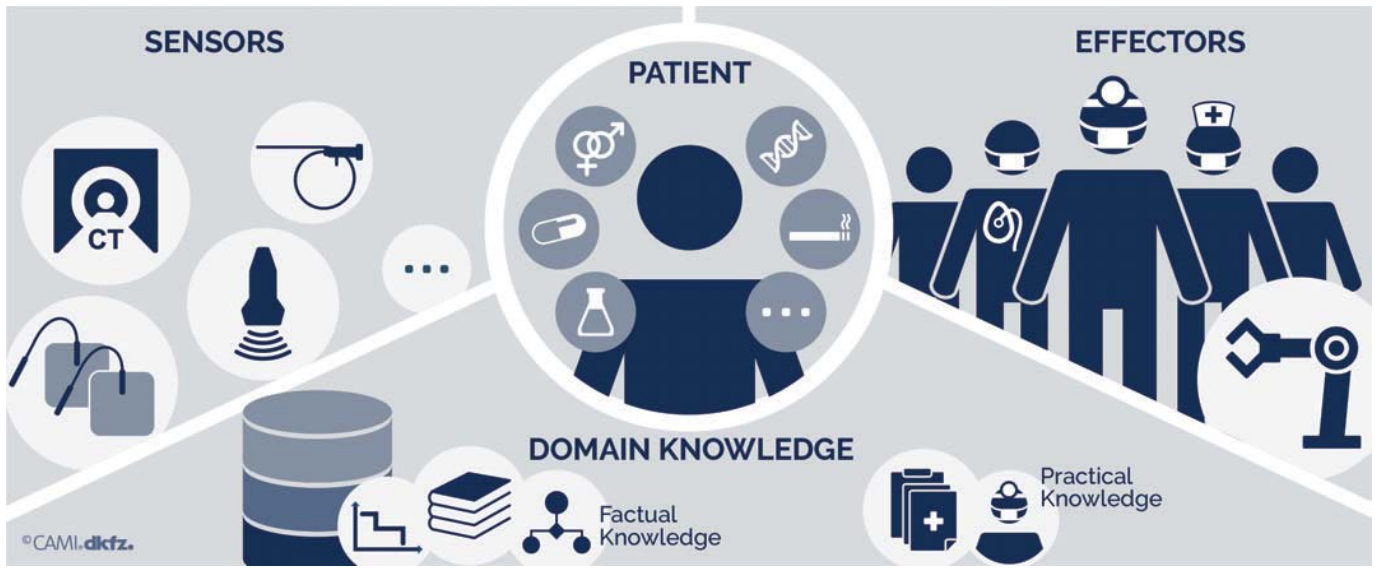

Fig. 2. Main components involved in SDS. Patient: The subject getting a diagnosis or treatment. Effectors: Humans and/or devices involved in the manipulation of the patient including surgeons, anesthesia team, nurses and robots. Sensors: Devices for perceiving patient- and procedure-related data such as images, vital signals and motion data from effectors. Domain knowledge: Factual knowledge, such as previous findings from studies, clinical guidelines or (hospital-specific) standards related to the clinical workflow; as well as practical knowledge from previous procedures.

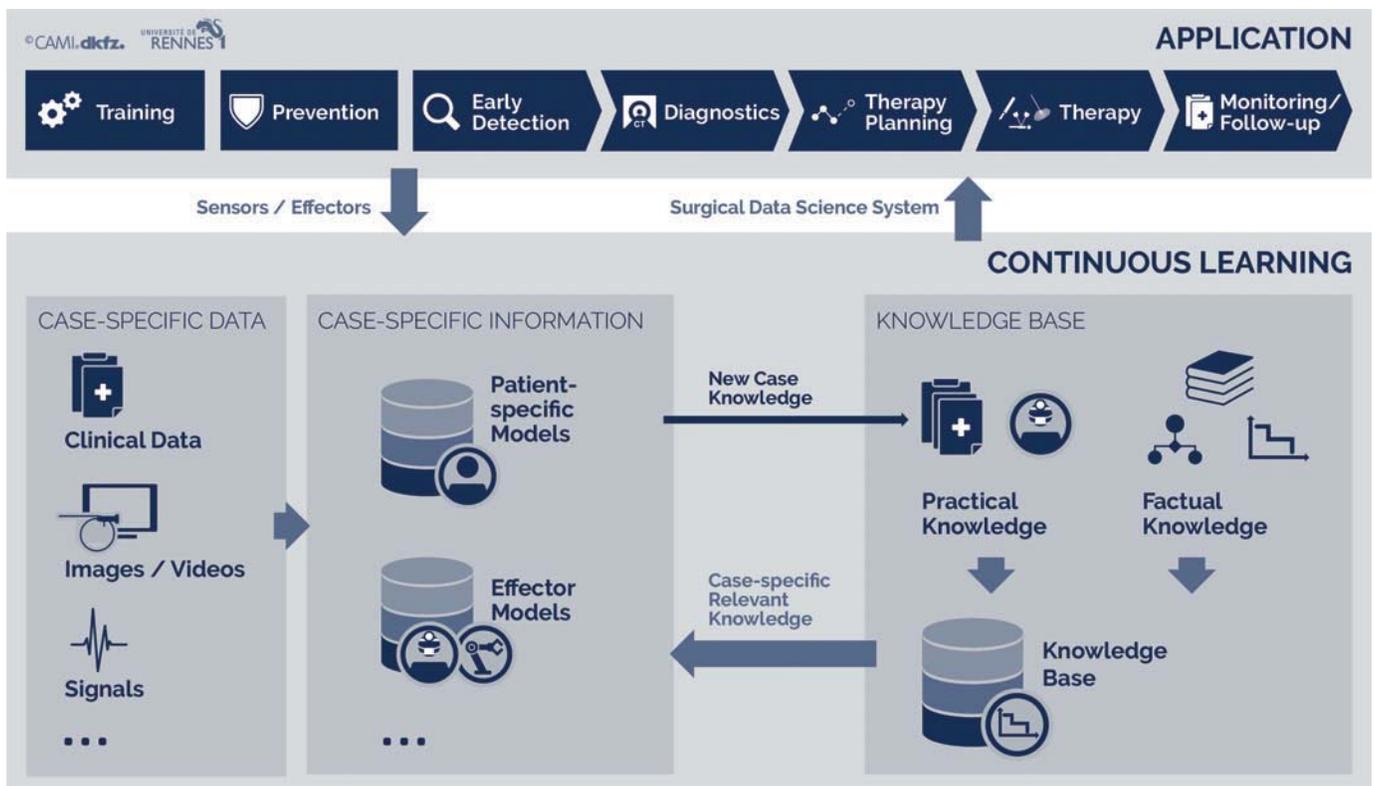

Fig. 3. Technical view on SDS.

The data and information flow in an SDS system can be summarized as follows:

1) Data acquisition: SDS relies on continuous acquisition of data about patient and caregivers during training, diagnosis, treatment planning, plan execution, assessment, and/or follow-up. In the case of surgical interventions, much of this data may be in the form of medical images taken before, during, or after the intervention but it may also include other clinical data such as lab results, past history and genomics data or vital signals acquired during an intervention.
2) Information generation: The perceived data is then further processed to obtain relevant information. For example, image processing methods may be applied to extract morphological or functional information from raw medical images, such as the vessel topology, the volume of an organ or information related to blood flow and perfusion. In this interpretation step, domain knowledge is typically used. For instance, active statistical shape models may be stored in the knowledge based in order to enable anatomical structure labeling in a knowledge-based manner. Also, case knowledge stored in the knowledge base may be used to derive an individual treatment decision for a patient.
3) Assistance: The information generated may then be used by an SDS system to provide context-aware assistance in various forms (Sec. III-B).
4) Update of the knowledge base: Finally, the new "case" is used to update the practical knowledge in the knowledge base.

*B. Range of clinical applications*
Authors: S. S. Vedula, H. Kenngott, M. Hashizume, R. Taylor

The narrative in this section is based upon responses from workshop participants for brainwriting question Q3 (Tab. II-A): "What do you consider to be key clinical applications of SDS?". The clinical applications identified by the participants broadly encompass training and patient care, and reflect both current research frontiers in these areas as well as a forward looking vision. Participants' responses to this question may be grouped into seven key clinical applications for SDS as shown in the upper part of Fig. 3. SDS can transform all aspects of the healthcare pathway, including prevention, early detection and diagnosis of disease, planning therapy and delivering it, and post-intervention care, in addition to training of healthcare providers.

*1) Training:* Successful surgical therapy requires skillful providers. Poor surgical technical skill is associated with an increased risk of severe adverse outcomes in patients, including readmission, reoperation, and death [7]. Thus, professional certification based on systematic skill assessment is necessary to ensure competent surgeons provide care. SDS can significantly support the training of skillful and competent surgeons. Much research in this regard currently is on data products for objective assessment of technical and non-technical skills, both for overall performance and for targeted segments of activity [8]. Granular assessment for activity segments enables directed feedback, which is necessary for skill acquisition through deliberate practice [9]. Granular assessment and data-driven feedback, in turn, rely upon techniques for automated detection of surgical activities [10]–[12]. Other modes of data-driven feedback include efficient retrieval of expert demonstrations of specific surgical activities. A data science approach to surgical training involves systematically capturing data on performance throughout training, from surgical rotations or clerkships as a medical student to graduate medical training. In this context, SDS can also provide tools to inform recruitment of medical students into surgical disciplines. Data products for skill assessment and directed feedback enable development of training curricula that aim to reduce time to competency [8]. An eventual goal for SDS in training is to support lifelong learning for practicing surgeons and continuous improvement in the quality and safety of care they provide to patients. This is possible through the use of analytics for skill and competency assessment via objective methods, using patient outcomes, as well as through development of an automated or virtual surgical coach [13].

*2) Prevention and early detection:* Preventive care and early detection of disease frequently includes a surgical intervention, e.g. endoscopy or internal tissue biopsy. While technology for preventive care has gradually evolved to enable minimally invasive access, SDS can potentially transform these technologies into data-driven knowledgeable systems. For example, colonoscopy is a routine intervention for early detection of colon cancer. Technology to capture motion of a colonoscope combined with techniques to register it in relation to polyps detected on a CT scan enables valuable navigation assistance to the surgeon [14]. SDS can also drive advances in detection of conditions that require surgical intervention upon diagnosis. For example, breast cancer is one of the most common cancers to be diagnosed across the world. Techniques to capture spatial and force patterns in how providers perform a clinical breast examination enables skill acquisition by providers through data-driven performance assessment and feedback [15]. While the preceding examples illustrate the application of data to surgical instruments and provider performance, wider adoption of wearables and other mobile technologies can serve as additional sources of rich data and drive future advances in prevention and early detection.

*3) Diagnostics:* Diagnostics in surgical care, particularly as it pertains to determining a surgical condition and whether to offer surgery for a patient, traditionally relied upon manual inspection of tissue or other discrete physiological or biochemical markers. SDS can transform surgical diagnostics through discovery of new data sources, enabling analysis of complex data sources, and advancing analytics to improve the accuracy of existing data sources. For example, histopathologic examination by an expert pathologist is integral to diagnosis of surgical diseases such as cancer. Image analytics to extract relevant features, and algorithms to detect cancer cells can enable not only automation but also improved accuracy in diagnosis [16]–[19]. Similar analyses of images, for example, from hyperspectral or multispectral imaging or confocal laser endomicroscopy, can assist with identifying cancer tissue during surgery [20]–[22]. Diagnostic surgical procedures can also see transformative changes through a data science approach. For example, techniques to automatically detect lesions such as polyps during colonoscopy can be useful to improve accuracy in the diagnosis of conditions that require surgical intervention [23], [24].

*4) Therapy planning:* Surgical intervention, in many instances, should be carefully tailored to the specific pathological and anatomical context in an individual to optimize benefit and minimize harm. Radiotherapy is a classic example where insufficient and inaccurate planning can lead to a therapy that is ineffective and harmful. Typically, providers rely upon preoperative imaging to develop a personalized plan to deliver a curative dose of radiation while minimizing potential toxicity. Numerous factors, such as anatomical information, patient demographics, intent of treatment, and radiation type and delivery, play a role in determining an optimal therapy plan for a given patient. The complexity of data for radiotherapy planning and the diversity in its sources necessitates automation to provide high quality patient care at scale. Recent literature illustrates the potential for machine learning techniques to develop radiotherapy plans [25], [26], and further advances in the development of scalable tools are possible through analytics that rely upon population-based atlases of anatomical contours and therapy plans [25], [27], [28]. Data-driven therapy planning can also play a critical role in care for patients undergoing other surgical procedures in sensitive anatomical areas, for example, surgery of the paranasal sinuses and skull base [29], and stereotactic brain surgery [30]. Finally, SDS enables further advances in planning surgical therapy through integration of various sources of information [6], including patient-specific data and simulations [31], [32], population-based atlases, and real-time data sources such as endoscopic video images [33].

*5) Therapy:* Effective surgical therapy involves appropriate manipulation of anatomical structures with adequate skill. SDS can drive future advances in delivery of surgical care by supporting decision-making and providers' skill. Large amounts of information are available for the surgeon to make decisions while delivering the intervention. SDS can potentially upend the current paradigm of decisions based upon sparse subsets of selective information to one based on advanced analytic models built using integrated data sources across large patient populations. Such advanced analytics can be supported with user-friendly interfaces to maximize value of information available to the surgeon while avoiding cognitive overload, i.e., provide surgeons with the right information at the right time. This encompasses intelligent data products that adapt in real-time to changes in operative strategy. Intelligent support during surgery may take other forms, such as vision-based guidance and navigation assistance. Data science tools to provide vision-based guidance during critical phases of the procedure may prevent surgical errors arising from visual misperceptions, for example, nerve injury during rectal cancer surgery [34], [35]. Similarly, navigation assistance is routinely employed in certain procedures such as endoscopic sinus surgery and ultrasound guided needle biopsies [36]. Intraoperative guidance and navigation assistance systems may further be adapted to specific procedures integrated with automated phase detection for safer and effective care. For example, tool guidance that adapts to shifts in tissue positions during surgery integrated with preoperative and intraoperative imaging may provide the least harmful and most effective way to reach lesions, which may then be resected or ablated. Similarly, automated localization of target anatomical structures can make surgery safer and prevent errors that are highly consequential for the patient [37]. Surgical patient care is a process. Surgical processes are complex owing to the intensity of care provided to patients and to variation across patients and caregivers. Data analytic tools to monitor surgical processes can improve safety and quality of patient care. Such tools rely upon data all along the patient care pathway, from initial presentation through surgical intervention to the end of follow-up. Within the operating room, surgical process monitoring can improve safety and quality of care through context awareness for automated systems. Context awareness involves continuously monitoring throughout the procedure patient status, evolution of surgical therapy as well as ancillary non-surgical interventions such as anesthesia, and intraoperative environmental factors. Typically, surgical procedures are a well-defined sequence of activities to manipulate anatomy as planned; they are also a sequential progression of interactions between operators and the environment. Monitoring surgical processes requires detecting which part of the procedure is being performed at any given time. Techniques for this purpose use models of either time series data from sensors or from human observations; details on such techniques are discussed elsewhere [38]. Surgical process models and techniques to detect surgical phases not only automate context awareness but also serve to characterize variation in processes and factors affecting it. Analytics on process variation may be used to improve efficiency of care, e.g. through standardization where possible, and detection of unanticipated deviations that influence patient outcomes [39], [40]. Surgical processes are complex. From the perspective of healthcare systems, the cross-sections of surgical patients at any point in time differ in their diagnoses, comorbidities, severity of illness, and planned interventions and post-intervention care. In addition to patient heterogeneity, complexity of surgical care processes is driven by several factors at different levels, such as a multitude of treatment options, stakeholders, and variation in facilities and techniques available to care for patients, and interactions between these factors. A data science approach is critical to discover inefficiencies, enhance transparency, and optimize complex surgical processes with respect to patient outcomes. SDS in this context aims to yield a deep understanding of care processes and patient outcomes, eventually leading to treatment plans individualized to patients and optimized for healthcare resources.

*6) Monitoring/follow-up:* SDS can have a substantial impact on how patients recover after surgical interventions in more than one way. Advances in data capture and analysis techniques can allow objective, data-driven measurements of outcomes at frequent timepoints that can be standardized across providers and institutions. In contrast, current approaches often rely upon unstandardized, subjective measures of outcomes assessed at sparse intervals in time. For example, surgical sites are routinely assessed for signs of infection through visual examination. In a data science approach, monitoring for surgical site infections involves an examination frequency adapted to the patient-specific risk identified through accurate predictive models and analytics applied to data sources such as images of the surgical site [41]–[44]. Another aspect of post-intervention follow-up care for patients that SDS can transform is prognosis. This is possible either through new methodologies applied to conventional data or discovery of new data sources. For example, systematic feature engineering techniques applied to patient registry based data can lead to discovery of new, clinically meaningful features to predict surgical outcomes [45]. Finally, SDS has a significant role in various aspects of postoperative care that affect eventual patient outcomes. This includes, for example, quantification of disability in patients resulting from surgery as well as recovery achieved with rehabilitation therapy. Thus, advances through SDS can potentially deliver new and improved clinical applications throughout the patient care pathway.

In summary, SDS can potentially advance surgical therapy through intelligent decision-support relying upon effective process monitoring, and through accurate assessment of providers' skill, efficient training, and automated coaching.

## IV. OPPORTUNITIES

### A. Key initiatives and projects
*Author: S. Speidel*

SDS is an emerging and challenging research field that includes different related research topics and questions. Following the brainwriting question "What do you consider to be the key initiatives/projects/standards in the field?" (Q7 in Tab. II-A) the workshop participants identified several projects which were classified as "Knowledge representation", "Data representation", "Data accessibility" or "Application perspectives". In addition a subsequent comprehensive survey of the state of the art was conducted and classified projects as either academic or commercial research. These categories were chosen since the identified large-scale projects cover several aspects of the above mentioned workshop classification.

*Academic Research:* An important aspect of data science in general are ontologies. These serve as a semantic representation of concepts and relations that are important in the application domain. Ontologies are a way of making domain knowledge explicit and machine readable, and as such serve as the basis for automatic information processing. The specific topic of ontologies for SDS is discussed further in Sec. VI-C. Several interdisciplinary collaborative large-scale research projects address different aspects of SDS. The Transregional Collaborative Research Center Cognition-Guided Surgery was developing a technical-cognitive assistance system for surgeons that explores new methods for knowledge-based decision support for surgery [6] as well as intraoperative assistance [46]. First steps towards the operating room (OR) of the future have been taken recently, focusing on different aspects like advanced imaging and robotics, multidimensional data modelling, acquisition and interpretation, as well as novel human-machine interfaces for a wide range of surgical and interventional applications (e.g. Advanced Multimodality Image Guided Operating [3], Computer Integrated Surgical Systems and Technology (CISST) Engineering Research Center [4], Hamlyn Centre [5], Wellcome Trust/EPSRC Centre for Surgical and Interventional Sciences [6], Innovation Center Computer Assisted Surgery [7], Institute of Image-guided Surgery [8], National Center for Tumor Diseases Dresden [9]. Furthermore, several multidisciplinary collaborative projects deal with surgical process analysis based on device signals from integrated sensor data in the OR. The Digital Operating Room Assistant DORA is a multidisciplinary project that focuses on optimizing the surgical process by monitoring and analyzing OR device systems [47]. SCOT (Smart Cyber Operating Theater) is a cooperation project for the operation room of the future in neurosurgery that analyzes all kinds of data including integrated devices [48] and visualizes relevant information for decision-making. The CONDOR project (Connected Optimized Network & Data in Operating Rooms) is another endeavor that aims to build a video-driven Surgical Control Tower within the new surgical facilities of the IHU Strasbourg hospital by developing a novel video standard and new surgical data analytics tools [10]. A similar initiative is The Operating Room of the Future (ORF) that does research on device integration in the OR, workflow process improvement, as well as decision support by combining patient data and OR devices for minimally invasive surgery [49]. Furthermore, a Surgical OR black box is proposed by Goldeberg et al. that enables synchronized data capture and analytics of multiple sensor feeds in the OR to prevent errors [50]. The former project OR.NET, that included several academic, clinical and industrial partners, now a non-profit organization OR.NET e.V., addresses the secure dynamic networking of components and transforms them into standardized activities which is a prerequisite for data analysis during surgery [11]. OR 4.1 is a German project funded by the Federal Ministry for Economic Affairs and Energy with the aim of developing a platform for the OR - in analogy to an operating system of smartphones - that allows for integration of new technical solutions via apps [12]. Technically, it builds upon several other projects including the OR.NET project and focuses on service-based business concepts. A primary goal is to facilitate clinical translation of research results.

*Commercial Research:* Commercial platforms and projects have so far focused mostly on analyzing multidimensional patient data for clinical decision-making and not on surgical applications. IBM Watson Health consists of several projects such as Watson Medical Sieve, Watson Oncology or Watson Clinical Matching which apply the Watson cognitive computing technology to different research questions in healthcare [51]. The goal of Watson Medical Sieve for example is to filter relevant information from patient records consisting of multimodal data to assist clinical decision making in radiology and cardiology. In addition Watson Clinical Matching finds clinical studies that match the conditions of individual patients. Google DeepMind Health uses their machine learning technology [52] to analyze medical data for diagnosis as well as providing time-sensitive information at the right time for physicians. Several pilot projects in radiotherapy, eye treatment and cancer therapy have been launched. Both Intel Healthcare Analytics [53] as well as Ascos CancerLinQ [54] make use of Big Data analytics to improve patient therapy for cancer. Intel Healthcare focuses on analyzing patient records with the processing architecture thereby reducing computational workload. CancerLinQ is a non-profit framework that uses the SAP Hana platform to extract data from multiple sources like electronic health records. Furthermore, hybrid as well as integrated operating rooms offered by several vendors like Siemens, Philips, GE, Toshiba, Storz or Stryker are evolving and offer infrastructure for SDS research.

---

[3] http://ncigt.org/amigo
[4] https://cisst.org/
[5] https://www.imperial.ac.uk/hamlyn-centre/
[6] http://www.ucl.ac.uk/interventional-surgical-sciences
[7] https://www.iccas.de/
[8] http://www.ihu-strasbourg.eu
[9] https://www.nct-dresden.de/en.html
[10] https://condor-project.eu/
[11] http://www.ornet.org
[12] http://www.op41.de/

In summary, there are many promising approaches which make use of innovative data science technologies. The academic research projects address several aspects of data science with a surgical application focus. In contrast the commercial platforms focus more on biomedical data science, analyzing patient data for decision-making in different clinical disciplines without a surgical application so far, except for vendors providing infrastructure for hybrid operating rooms.

### B. Related research fields
*Author: S. Giannarou*

SDS is a highly interdisciplinary research area, depending on and being closely related to various research fields that are either associated with the generation and processing of surgical data or involve application areas which can employ and benefit from the tools produced from SDS. The present section highlights some of the most important related research fields based on the brainwriting question "Q8: What are related domains/research fields?" (Tab. II-A), namely robotics, machine learning, omics, and statistics.

One of the main research fields which is closely related to SDS is robotics. Recent advances in surgical robots for Minimally Invasive Surgery (MIS) have allowed the recording of intraoperative video as well as kinematic data, including position of tools, angles between robot joints, velocity and force/torque measurements. The processing of this information has enabled the development of "intelligent" machines which can extend the surgeon's capacities, make decisions and independently execute surgical tasks. For instance, in robot-assisted MIS, the integration of vision and kinematic data has enabled the autonomous execution of common, repetitive and well-defined tasks which might be ergonomically difficult for the surgeon. To this end, current research has mainly focused on robot-assisted ultrasound elastography [55], motion compensation in cardiovascular surgery [56], autonomous tissue dissection [57], brain ablation [58], and endomicroscopy scanning [59] under image guidance. Autonomous execution of surgical tasks that require repetitive and precise motion can significantly reduce the cognitive load of the surgeon during the operation. These intelligent robotic platforms can also filter out hand tremor and allow dexterous maneuvers, improving outcome and minimizing trauma to the patient.

For the processing of the above information, machine learning has been extensively used in SDS to analyze pre- and intraoperative data. Medical image segmentation and registration are two of the most representative applications. The former aims at delineating the borders of important anatomical structures or regions of interest, while the latter brings structures of interest on separate images into spatial alignment [60], [61]. The registration of preoperatively segmented structures to intraoperative data can guide the surgeon in the Operating Room (OR) to target previously identified pathological areas. Machine learning has also been integrated into Computer Vision techniques to model the context of the surgical environment and the activity in the OR. The representation and analysis of the context of the surgical environment involves processing video data to track surgical tools [62], [63], monitoring their usage and modeling their motion as well as the tool-tissue interaction [64]. To model the activity in the OR, vision-based approaches have been proposed to detect and track the OR staff. For this purpose, multiple cameras have been installed in the OR to capture colour and depth images to detect the surgical staff and estimate their pose [65]. These vision-based approaches have the advantage that they do not suffer from the limitations of body-worn sensors tracking systems which are invasive and difficult to introduce in the OR.

Biology and in particular the "OMICS" field is another large-scale data-rich research area that can provide valuable insight for SDS. Data from this field aims at the collective characterization of the roles, relationships, and actions of the various types of molecules that make up the cells of an organism. This information is important for diagnosis and treatment planning. To deal with the big volume of fragmented clinical data generated from pre-, intra- and postoperative tasks, the Semantic Web has been recently employed in clinical applications to establish a common framework for data representation [66]. The Semantic Web uses semantic languages and tools for ontologies and metadata management to provide machine-readable information [67]. This allows computers to automatically organize the content of information spread across multiple pages, interpret information and perform tasks on behalf on the clinicians. Establishing a well-defined, standardized knowledge representation, the Semantic Web facilitates the integration of clinical information and allows data to be shared and reused across organizations. The topic of ontologies is discussed in more detail in sec. VI-C.

Statistical analysis is another powerful processing tool that enables clinicians to draw meaningful conclusions from medical data collected through observation or experimentation. This might involve comparing different treatments, assessing the effectiveness of a medication, diagnosing the type of cancer or monitoring the health of patients after treatment. Raw biomedical signals, "OMICS" signatures, demographic data, medical reports and others can provide input data for statistical analysis in order to extract characteristics which can assist diagnosis and support decision making.

In turn, other research fields will be able to benefit from the developments in SDS. Medical robot control, for example, can introduce novel models for human-robot interaction towards the design of cognitive humanoid robots. Surgical vision for intraoperative navigation within complex and deformable environments will advance computer vision techniques for a wide range of applications, such as industrial inspection. Also, the research on automatic surgical task execution could assist autonomous vehicle driving. The developments on intraoperative augmented reality visualization can be used in multimedia applications and game platforms. The biomechanical tissue modeling and anatomical constraint models can advance new haptic interfaces. The real-time performance requirement of intraoperative SDS data processing could inspire high performance processing systems, as well as the design of efficient engines for the management and processing of "big data".

While the majority of the above SDS systems are not ready yet for use on humans, they do represent the future of surgery. Technological advances are rapidly being developed in academic and industrial labs but extensive validation and trials on mock-ups and ultimately human subjects is required. A detailed analysis on the challenges that need to be addressed to widely establish the use of SDS in the OR can be found in the following sections.

## V. Risks and Challenges

### A. Technical risks and challenges
*Author: N. Padoy*

The present section highlights the most important technical risks and challenges as identified by the workshop participants based on the brainwriting question "Q5: What do you regard as key technical challenges in surgical data science?" (Tab. II-A).

As described in Section III-A, a key aspect of SDS is the inclusion of procedural data. SDS pertains to all the processes taking place within the operating room and is highly related to pre- and post- operative processes as well. A fundamental assumption of SDS is therefore that these processes can all be perceived digitally to enable the computational processing required for data analysis and related applications. We identify four main groups of challenges related to this automated computational processing:

*1) OR infrastructure:* Even though the OR is becoming increasingly digital, surgical devices generally use proprietary interfaces that do not allow the straightforward acquisition of the information they process or present to the clinical staff [68]. To deliver the promises of SDS, it is necessary that all devices, equipment and information systems become integrated into a common data acquisition framework allowing further processing. This will require standardization across various suppliers, as already taken into consideration within the OR.NET efforts [69]. In this endeavor, it will be crucial that the devices have access to a common OR clock to ensure that the data can be accessed with accurate timestamps and also synchronously when needed. Such an acquisition framework should be open to the inclusion of new sensors and systems as these get introduced to the OR. To handle and store the large amount of new digital data, the OR will also need a central repository that will complement or include the actual PACS system. Since the effectiveness of SDS will be strongly increased by the use of data from multiple institutions, the system should provide ways to anonymize and share this data and the models generated from this data. Standards and protocols will be crucial to avoid masses of data being acquired by custom systems that are not inter-operable [70].

*2) Real-time data processing:* It is likely that the multi-modal data used in processing pipelines will be highly heterogeneous as they will not only come from equipment from different providers, but also from potentially very different equipment depending on the type of surgery, surgeon preferences, operating room, hospital and country. Furthermore, the data will be highly multi-modal, ranging from medical images and discrete events to temporal streams of patient vitals and videos. The methods for data analysis, modelling and processing will need to accommodate this heterogeneity, which comes in addition to possible variations in the device configuration parameters and calibrations [31], [32]. Furthermore, the developed methods will need to adapt to the heterogeneity of the processes themselves, as the same kind of procedures may be performed differently by teams composed of varying numbers of actors with different habits and skills. It will be particularly important to make sure that the models used to process the data have sufficient generalization for them to be used in more surgical contexts than one particular OR. It will also be important to ensure that all surgeries benefit from SDS, including less common surgeries that generate less data. In this latter case, cooperation across institutions will be essential.

*3) Smart user interfaces:* Ultimately, SDS should result in applications and tools used during and around the surgery for assistance, decision support and process analysis. This will require user interfaces that present this information and associated interactions in a summarized and human understandable manner [71]. Intra-operatively, the OR data should also be used to recognize the current context [12] and present the information in a context-sensitive manner. It is indeed important that these tools become quickly accepted and not left out in a corner of the OR. Focus should be on ease-of-use and comfortable context-aware presentation and interactions, since one objective of data-science is to improve workflows by using the wealth of available OR data while at the same time avoiding data overload to clinicians and staff.

*4) Access to data and tools:* The real benefits of SDS will appear through the democratization of its applications and analyses within different institutions. This will require dedicated collaborative tools (such as Protege [72] for ontology engineering), semantic knowledge data bases and shared platforms (such as described in [73]) and open source software (such as the Medical Simulation Markup Language [74]) accepted by a wide range of engineers, clinicians and staff. These tools should allow the validation of the findings through replication in other hospitals, also in situations where the input data contains variations or is incomplete/different from the data available in the institution at the origin of the study. Consequently, the community should also work on the development of common benchmarks for SDS approaches that are robust to the inevitable data heterogeneity. This aspect is discussed in more detail in Sec. VI-D.

*B. Moral, ethical and social challenges*
Author: T. Neumuth

The present section highlights the most important challenges as identified by the workshop participants based on the brainwriting question "Q6: What do you consider to be moral, ethical or social challenges related to SDS?" (Tab. II-A).

The rise of SDS is a novel socio-technical phenomenon: Recent technical advances enable researchers to deduce more correlations and patterns from gathered data than ever before [75]. These deductions allow for predictions that were previously impossible due to the lack of information. SDS has the potential to facilitate a new system of interventional knowledge. It changes the knowledge objectives and enables new approaches to understanding information entities and their interrelations. Furthermore, it proposes new key questions about the constitution of intervention-related knowledge. Finally, there is a chance of automating knowledge constitution.

The quantification of individual actions, sensory data, and other real world measurements has the power to create a digital image of interventions. Such images constitute the basis of SDS. The analysis of digital images extends the scope of surgical research in general. Today, however, there is no culture of collecting large data volumes in surgery. A culture change towards evidence and evaluation supported by SDS is necessary. For this aim, it is urgent to gain the support of both surgeons and the surgical community.

Additionally, these new approaches and possibilities raise many societal, ethical, and moral challenges. Will surgical data analysis help create better interventional approaches, tools, or services? Will it enhance our understanding of interventional rationales? And how to engage non-technical users for general support of the approach?

SDS can be a powerful tool to support the tackling of recent societal challenges. The aging population or cancer research can benefit from therapy optimization, to give just one example. Stakeholders could cooperate closely to extract potentialities from the data generated by patients, clinicians, or technology. These entities put out a great amount of data as result of their being, circumstances, doing, or functioning. Data collectors collect and store data, and govern collection and utility. Data utilizers from research or industry (re-)define the purpose for which data is used. They lay down new knowledge rules or create innovation by combining data sets.

The rising approach might create power imbalances between data stakeholders [76]. The stakeholders that produce data and those that generate intelligence from the data are separate entities. Yet, missing knowledge or understanding of data generators, especially patients, about which data is collected or what it is used for puts them at an ethical disadvantage. To avoid this situation, moral agency, the moral responsibility of the other stakeholders, comes into play. The ethics of SDS might shift control from a personal moral agency and increase moral liability of those that have control over the data. Data collection also creates new gaps. Collecting surgical data is laborious, time consuming, and resource intensive.

And yet many questions still remain unanswered. Who gets access to which data? What are the purposes and what the constraints? There is a risk that new hierarchies might arise, ranging from stakeholders that are surgical data "rich" to those who are surgical data "poor". This situation would clearly put researchers from "poor" institutions at great disadvantage.

But even if the "rich" were willing to share their data there is still the issue of patient consent. Breaches in privacy are hard to detect and to label as such. Which strategies are on hand to prevent data triangulation for deanonymizing patients [77]? Is it appropriate, feasible, and proportionate to get consent from every patient? To what extend will patient data be accessible for research? Can researchers use some data without requesting permission? Data regulations are only made with regard to current research. How can researchers demand that patients give their consent to the use of their data for research that is not even thought of at that point of time?

Nowadays, more data is available and accessible than ever before in the history of surgery. No previous experience in handling the data flood is available and inbuilt flaws have not yet surfaced [78]. The most sensible levels of data abstraction, generalization, or patient specificity are yet unknown and need to be defined.

Just concentrating on analyzing the data does not support science. It is problematic for researchers to justify data analyses just because the data is accessible or available. This holds especially true for data provided by artificial actors such as computerized medical devices. Data analysis comprises the risk that it emphasizes correlation over causation [76]. This phenomenon, referred to as apophenia, tends to see patterns where none exist. It stems from the issue that more data has more interconnections. Since there is limited experience in interpreting large volume data sets, interpreting results without coherent models or unified theories must be realized with caution [79]. Reliability of automation requires validation methods to prevent system overtrust; new research processes are required and the way information is used needs to be clarified [78]. We need to bear in mind that research insights are discoverable at any level; focusing on an individual can also provide important insights. The data volume should be appropriate to the research question asked and responsibilities for malfunctions of the systems need to be resolved.

The advent of SDS will change the performance of research; our current decisions will have an impact on the future. Surely, SDS will be judged by its impact on patient outcome, it will change knowledge and enable new approaches to understanding information entities and their interrelations. In short: it will affect the future of surgical and therefore medical research in general.

## VI. ROADMAP

To propose a roadmap for advancing the field of SDS, the following aspects were discussed: What are the key research questions to be addressed (Sec. VI-A)? What are the key clinical applications to focus on (Sec. VI-B)? How to work towards a shared ontology (Sec. VI-C), shared data repositories (Sec. VI-D) and shared tools (sec. VI-E)? How to generate incentives for advancing the field (Sec. VI-G)? How to disseminate SDS methods (Sec. VI-H)?

*A. Key research questions*
Authors: N. Navab and R. Taylor

The current section is based on World Café Question W1: "Key research questions: What are the key (highest priority) research questions?" (Tab. II-B). Even if the term SDS and the idea of creating a scientific community around this theme is novel, different components of this research have been the focus of many scientists in the last few decades. Heinz Lemke is one of the main scientists who have been advocating the need for research on many components of what we now call SDS for the last few decades [80]–[82]. However, because of the particular characteristics of surgery and its challenging requirements compared to diagnostics, one had to wait for a few decades before the availability of high computational power and advances in different fields of computational science would allow the community to define SDS as a novel focus area with its particular characteristics. Let us take a more precise look at the particular characteristics of surgery and its specific requirements [83]:

1) Relevance: compared to the diagnostic process, in which the acquired data are first of general nature and only gradually become specific to a given disease or anatomy, the surgical process requires the acquisition of particular relevant information for a given procedure often designed for a specific patient.
2) Speed: as soon as the surgical procedure starts, the whole crew is under considerable time pressure and often time is life. This is not the case in diagnostic processes and dramatically changes the nature of research on information processing in the respective fields of diagnosis and surgery.
3) Flexibility: while the diagnostic process follows a rather regular flow of data acquisition, the surgical process varies significantly and is highly process and patient specific.
4) Collaboration: the surgical crew often acts as one single unit. While the main surgeon has the lead, anesthetists, assistant surgeons, circulators, nurses and staff play crucial roles at different task flow steps within surgery. Their smooth, dynamic collaboration and coordination is a crucial factor for the success of the overall process.
5) User interface: because of all the above requirements, user interfaces are complex to design as heterogeneous information is needed throughout the process. Different players within the surgical suite may require the same or different information simultaneously or at different times. The number of user interfaces in operating rooms is continuously increasing as novel imaging, instrumentation and patient monitoring tools enter the operating rooms.
6) User interaction: because of sterility issues and the need for dynamic information during surgical acts, the interaction with user interfaces is complex and is a permanent issue to be dealt with for some of the same reasons discussed above.
7) Safety and ethics: the case of safety is also complicated within a surgical process. Surgeons and surgical crew need to decide dynamically for example on the amount of radiation exposure and risk taking. This is always an ethical and complex decision when the life of the patient is at the stake.

Surgeons have historically acted as the head of an orchestra, deciding on the surgical workflow, defining the instruments and coordinating the crew to optimize the overall outcome of the surgery. Each provider has focused on producing one or multiple sets of devices and different surgeons have decided to opt for different tools and technologies. The surgical environment has therefore integrated many solutions provided by a large variety of providers, making the unification and analysis of dynamic information difficult and only possible for intelligent and well trained surgeons and surgical crews.

The principle of 'see-one, do-one and teach-one' [84] has also made the integration of novel technology into surgical environments complex and the evolution of surgical processes often lengthy and difficult. Critically novel surgical methods such as minimally invasive surgery introduced a few decades ago [85], and robotic surgery at the beginning of this century [86], have enabled radical changes in such environments. Changing the basic infra-structure and information flow in traditional surgeries remains harder than introduction of novel surgery techniques because the former requires redesign of the whole surgical environment in order to integrate the digitalization and display required for data gathering, fusion and representation.

The described requirements, characteristics and historical development of surgical fields requires SDS to encompass many aspects of research and to embrace a set of heterogeneous research directions and activities. Some of these SDS research sub-fields, each defining their own specific challenges, are:

1) Advanced patient digital data acquisition, anonymization, storage and handling. This challenge is discussed in more detail in the workshop *Comment* [3].
2) Surgical process data acquisition:
    a) Digital data handling, communication and management within surgical theatres.
    b) Advanced sensing, including but not limited to computer vision 2D and 3D technologies [87], audio sensing [88], biophotonics techniques [89], [90] and other sensory, tracking and identification technology [91].
3) Large-scale data annotation: Methods to facilitate large-scale data annotation, e.g. based on concepts of crowdsourcing [92], [93], expert data augmentation [94] or self-supervised learning [95].
4) Ethical and social studies of medical data acquisition, storage and handling.

5) Data analytics:
   a) Applied to a large variety set of heterogeneous data including genetics, biomarkers, patient information, imaging, pre- and intra-operative data, enabling the move from evidence-based to knowledge-based medicine [6].
   b) Applied to surgical planning, execution and outcome, allowing better understanding of surgery and full analysis of existing techniques, their advantages and shortcomings.
6) Surgical process modeling:
   a) For understanding, modeling, classification, analysis and design, in which case it does not need to be based on on-line digital data acquisition [91], [96]
   b) For real-time monitoring of the surgical process, requiring real-time digital data acquisition and processing.
7) User interface: research on user interface design is of particular importance as it not only influences the SDS but also gets influenced by SDS [97].
8) Surgical skill assessment: as one of the main objectives of SDS is the improvement of surgical performance, assessment of surgical skills allows for indirect assessment of the improvement SDS brings in, on the other hand methods and technologies for evaluating the surgical skills are often also useful for modeling and monitoring of surgical procedures. It is important to notice that surgical skill assessment is also a large field of research including:
   a) Analysis of surgical dexterity: different research groups have been developing surgical dexterity measurement phantoms, tools and methodologies. In addition, access to kinematics data from robots used in surgical procedures has allowed such fields to generate valuable data and make considerable progress.
   b) Analysis of cognitive skills in surgery: this is much more challenging and harder for the scientific community to propose automatic systems and methods for.
   c) Analysis of communication and interaction in surgical theatres: a few groups have proposed to develop full environment surgery simulation [98], [99] moving towards analysis of the overall surgical performance not only in terms of dexterity and cognition, but also in terms of team communication and collaboration, throughout high intensity and complex procedures.

The above does not present an exhaustive set of research fields. Science and technology constantly advances and novel procedures and surgical techniques are regularly introduced. These may bring additional research field which need to be considered soon as part of SDS. This section only focuses on what the authors consider as research subjects which need to be addressed in the near future and will attract researchers from different fields to join the young but growing SDS community.

*B. Key clinical applications short/mid/long-term*
Authors: M. Kranzfelder, H. Feussner and B. Müller

Surgery of today – and even more in the future – is situated in a highly competitive environment among other interventional medical disciplines. Many former "surgical diseases" are now treated successfully either conservatively (e.g. peptic ulcers) or by interventional gastroenterology (e.g. endoscopic submucosal dissection for early esophageal cancer stages) or radiology (e.g. ablation of liver tumors). Accordingly, the field of surgery has to improve continuously if it is willing to maintain its key role in interventional medicine. To do so, improvement and optimization of the surgical discipline has to have at least three dimensions: a better therapeutic outcome, a reduction of invasiveness and trauma, and simultaneously a higher efficiency, i.e. sparing of resources.

Although these different aims seem to be incompatible at first glance, there is one possibility to combine them: by the consequent use of technical innovations and knowledge. First, a highly individualized preoperative therapy planning including sophisticated diagnostic workup (comprehensive preoperative imaging), tailored surgical simulation (evaluation of extent of surgical procedure) and proper resource assessment (in/outpatient treatment) are needed. Second, use of advanced surgical instruments, that are to some extent adaptive and cooperative ("intelligent" instruments, e.g. shape adjustment after insertion into abdominal cavity) should enhance safety and surgical performance intraoperatively. Last (but not least), an optimized postoperative care should increase patient comfort and safety by reduction of postoperative pain, length of hospital stay and – of utmost importance – morbidity and mortality to the minimum.

The main objective of SDS is the enhancement of surgical care by means of acquisition, modeling, interpretation and analysis of data that is/can be obtained pre-, intra- and postoperatively. It could become the key to integrate all modules and dimensions of modern surgery, e.g. optimization of the clinical workflow process (including the operation as a small part), improved imaging, advanced visualization, integration of robotics, etc. into a new comprehensive approach to surgery: cognitive surgery. There is almost no aspect of current or future surgery that is able to develop without the active support of SDS. The only question is which application will profit most and in which temporal order. This section therefore discusses the World Café Question W2: "What are the high impact clinical applications short-term, mid-term, long-term?" (Tab. II-B).

*1) Short-term applications:* In general, short-term applications should include direct and easy access to a comprehensive data pool of information on a surgical/medical procedure, classification of information by means of relevance and accuracy and the integration of this information into appropriate software platforms.

Thus, a precondition for the use of SDS is that digitalized information is available and can be provided. This holds true already for preoperative diagnostics today. Laboratory findings, functional examinations and, in particular, diagnostic imaging already enable the creation of a comprehensive "patient model" [6]. Based upon the individualized patient model, a precise and reliable therapy planning should become feasible in the short-term. In addition, as soon as the parameters of surgical decision-making are well defined based upon the patient model, a bench marking in the framework of surgical education and training is conceivable. Thus, surgical education can be made more effective.

Another short-term impact of SDS can be expected from the improvement of man-machine-interfaces based upon speech interpretation and recognition systems. This is mainly driven by the progress in the consumer market. Advanced systems such as SIRI (Apple Inc., Cupertino, California, USA) or ALEXA (Amazon.com, Inc., Seattle, Washington, USA) now offer extension tools for user specific applications. It is just a matter of a short time that applications for the surgical domain, e.g. dedicated speech control interfaces for robotic systems will be developed – either by ourselves, or, if we hesitate in seizing the opportunity – by the industrial companies. The same holds true for novel gesture based systems such as the Wii system (Nintendo, Kyoto, Japan), the Leap Motion system (Leap Motion Inc., San Francisco, USA) or MYO wristbands (Thalmic Labs Inc., Waterloo, Canada), which could track hand movements to interact with the integrated operating room suite to browse clinical data or to adjust settings of medical devices.

*2) Mid-term applications:* The modern surgical OR comprises of a variety of different technical units and devices that are deployed dependent on the course of a procedure to a variable extent. As the technical requirements subsequently increase, the surgical team would certainly benefit from a communication system in colloquial language with the surrounding technical OR environment. Today, the surgeon's intention to adjust the technical periphery according to his or her needs (e.g. light status on/off, tilt OR table, call for the next patient, etc.) is "translated" by a human assistant, most often the circulator. Not infrequently, this makes the process slow and prone to failures. A direct and reliable interaction between the surgeon and the high-tech OR environment would therefore be highly valuable.

One may assume that identification and recognition interfaces for tracking and tracing of objects/persons in the OR will become increasingly cognitive with the advancement of SDS, since they could be embedded within a techno-surgical environment that "understands" the course of the procedure and the potentially necessary steps in each particular situation. However, this can only be expected in the mid-term, since the main requirements – e.g. comprehensive, real time data capturing and a highly granulated modelling of the surgical procedure still have to be created. However, at the end of this process, the surgical OR will have changed from the OR of today with its agglomeration of dedicated technical units, devices and systems to a fully integrated, cooperative functional unit.

The use of virtual reality for preoperative training and operation planning as well as augmented reality inside the operating room during the procedure will further enhance accessibility of relevant preoperative data at point-of-care and will be compatible with a sterile environment. This will lead to improved understanding of patient cases, minimizing errors due to missing information, potentially accelerating decision making and procedure performance as well as improving training and skill assessment of young surgeons.

Another mid-term application is the systematic utilization of the vast amount of clinical data that are generated during each individual patient treatment, including the preoperative diagnostic workup, the type and course of treatment, and the postoperative outcome [6]. If these data are collected systematically in a well-defined structure, evidence could be extracted to facilitate further evidence based clinical decision-making.

This idea of "data mining" is not at all new and has already been proven to be helpful. More than 30 years ago, the Japanese surgeon K. Maruyama generated a prospective database of almost 1,931 gastric cancer patients mostly manually that enabled him to predict precisely the inflicted lymph node groups and thereby allowed for a "tailored resection" and survival rate estimation in gastric cancer for the first time [100]. The validity of this first approach of knowledge extraction was confirmed by a study in Europe [101].

Nonetheless, this idea of K. Maruyama did not become widely accepted, certainly because it was too far ahead of its time. Furthermore, manual data collection and documentation was by far too time consuming and too work intensive within the framework of routine clinical care. However, with the consistent progress in computing power and new approaches to large-scale data annotation [92] the necessary precondition, e.g. automated data retrieval, can be met more easily today. Accordingly, we can assume that surgical decision-making can be based, in the mid-term, on scientific data derived from statistical evidence elaborated from hundreds or thousands of similar cases, which have been treated before all over the world.

*3) Long-term applications:* Long-term applications mainly aim towards processing of medical devices and optimization of surgical workflow. The latter includes development and application of cooperative and adaptive (semi) autonomous systems that should facilitate individualized targeted therapy and diagnostics.

Surgery itself, i.e. the surgical manipulation during the operation, will draw a significant benefit from SDS. Many surgeons do accept that digitalized surgery and SDS will play an increasingly important role in preoperative therapy planning, etc., but not only a few are firmly convinced that the real act of surgery will remain an art rather than become a science. Accordingly, doubts exist whether SDS is really appropriate to become the scientific key to all aspects of modern surgery.

The increasing clinical use of mechatronic support systems (so-called "surgical robots", e.g. the da Vinci® surgical system, Intuitive Surgical Inc., Sunnyvale, California, USA) is a striking argument favoring the prospect that SDS already affects the surgical environment directly in the OR today. Although the da Vinci® and the new competitors such as (SPORT®, Titan Medical, Ontario, Canada; Senhance®, TransEnterix Inc., Morrisville, NC, USA) are basically nothing more than advanced tele-manipulating systems, their increased deployment is an indicator for the future development of surgical procedures. Evidently, at least parts of an operation will no longer be performed directly by the human hand in the future, but by a machine under human control. To increase effectiveness and, in addition, perform operations at a higher safety level than current conventional surgical procedures, SDS is indispensable.

In addition, the OR itself will be bound to a further transformation by digitization process. Future ORs will mostly be characterized by their integrated software suites. Those systems will allow for at least semi-automatic aggregation of clinical data and will have workflow and clinical process understanding. This again allows for the developing of cognitive abilities and providing situation-awareness thus acting similarly to human assistants throughout procedures using intuitive man-machine interfaces. Ultimately we may perceive the OR not as a room but as a truly human-like assistant.

It is a future vision of both surgeons and computer scientists to perform surgery in a fully integrated, cognitive and cooperative environment supported by (semi)autonomous robotic systems. All of the latter are integrated into the internet of things (IoT) enabling them to communicate with each other with simple and less critical decisions made on the machine level.

Step by step, surgery will undergo the difficult metamorphosis from an art to science. To some, this might be regretful, but the development is inevitable. However, this also opens up new opportunities. Up to now, the surgical discipline always had difficulties to cultivate its own knowledge domain (surgical science), since surgical clinical research usually takes a long time and is costly. Accordingly, young surgeons rather prefer laboratory studies which can be finished in a reasonable period of time. However, they are often not very close to real surgical topics [102]. SDS could offer a realistic new option for young surgeons with academic ambitions: It not only provides a faster path to earn scientific merits but also adds relevant new knowledge to the surgical domain.

*C. Towards a shared ontology*
Author: G. Forestier

The current section is based on World Café Question W3: "Shared ontologies: Do we need a common ontology? What do we have? How do we proceed?" (Tab. II-B).

Similarly to data science, SDS faces the challenge of rapid growth of the amount of recorded data. In this context, there is an urgent need for methods capable of exploiting these data and making sense of them. Thus, it is important that researchers and industry have access to a common resource describing precisely the types, properties, and interrelationships of the main concepts of this field. This modeling effort is mandatory to capture the always growing landscape of SDS and to ensure a proper definition of this field. In this context, ontology engineering offers an elegant solution by providing modeling methodologies and software to represent, define and manipulate the concepts of a domain. An ontology, in its simplest form, can be seen as a controlled and structured vocabulary of general terms that represent the main entities of a domain. These terms can be organized into a hierarchy using subtype relation to create a taxonomy providing different granularity levels of the domain. Going beyond the list of terms, complex relations between the entities and logical definitions can be defined to represent domain knowledge, thus enabling semantic querying and reasoning, thanks to reasoning engines and semantic query systems. Standard languages such as the Research Description Framework (RDF), RDFS and the Web Ontology Language (OWL) [103] have been defined by the World Wide Web Consortium (W3C), that guarantee interoperability between ontology resources and datasets based on these ontologies.

The primary use of ontologies is the annotation or tagging of resources (text documents, images, processes, etc). For example, impressive results have been obtained in the field of organism biology with the human genome project [104] and the development of the Gene Ontology (GO), intensively used for annotating experimental data and literature. The GO is presented as a tool for the unification of biology, illustrating the federating power of the construction of a central ontology for a research field. In biomedical imaging, ontologies have also been successfully used to promote interoperability of heterogeneous data through consistent tagging [105], [106]. The Foundational Model of Anatomy (FMA) [107] is also an important resource aiming at comprehensively modeling the anatomy of the human body. Furthermore, SNOMED CT [108] or Radlex [109] are also relevant ontologies with potential use in surgery. Existing initiatives like the Open Biological and Biomedical Ontologies (OBO) Foundry [110] project, that focuses on biology and biomedicine, also show that building and sharing inter-operable ontologies stimulate data sharing in a domain. All ontologies from the OBO-Foundry share the Basic Formal Ontology (BFO) [111] as a common top-level ontology and common design principles to ensure interoperability. However, somewhat surprisingly, little attention has been put to development of shared resources in the context of surgery.

The OntoSPM [112] project is the first initiative whose goal is the modeling of the entities of surgical process models, of which LapOntoSPM [113] is a derivation for laparoscopic surgery. OntoSPM is now organized as a collaborative action associating a dozen research institutions in Europe, with the primary goal of specifying a core ontology of surgical processes, thus gathering the basic vocabulary to describe surgical actions, instruments, actors, and their roles. A draft ontology of this core ontology exists, but needs to be extended to cover other related domains such as pre and intra-operative imaging, robotic surgery and surgery simulation, for which ontologies are already available. [114]–[116]. The strategy is to align and integrate (at least in part) such ontologies, rather than re-develop them. Adoption by the community of SDS is a critical challenge, and this is the reason why clinicians and researchers should be associated very early in this effort, together with software engineers and experts of the multiple application fields involved (see Sec. VI-B). The definition of the field of SDS as presented in this paper is the first step towards a consensus on the domain such an ontology should ultimately cover.

Once available and broadly adopted, a shared ontology would stimulate the community and boost data and knowledge exchange in the whole domain of SDS. In detail, the benefits of a shared ontology include workflow annotation, data annotation, data sharing, and improvement of interoperability between data repositories and systems. Beyond this direct added value in research, shared ontologies would greatly facilitate the development of international standards for integrated OR (interfacing of sensors, effectors, reporting systems, etc.).

*D. Towards shared data repositories*
Author: G. Hager, S. Speidel

The current section is based on World Café Question W4: "Shared data access: How can we create an international clinical data repository?" (Tab. II-B).

Progress in SDS will depend crucially on collaborative efforts to create data resources and establish methodologies that will underpin fundamental advances in the field. To date, there are a handful of shared data sets available to the field. However, these data sets are small, they are often tied to a single institution, and they are extremely diverse in structure, nomenclature, and target procedure (Tab. VI-D).

Future data sets will become the basis by which the field measures progress. Some of the attributes data sets must have include the following:

1) When taken in aggregate, they should establish a link from surgical training, to surgical performance, and surgical outcomes.
2) They must span multiple institutions using an agreed upon set of protocols and conventions. Documentation and infrastructure to augment data sets should be included.
3) They must be at a scale where statistical significance can be ensured even as multiple groups develop and test against them.
4) Be combined with well-defined criteria for validation and replication of results so that results can be compared across different groups.

Achieving these goals requires the field to establish the technical means to collect and share data, adopt methodologies and tools to support reproducible science [117], and to create a culture of data and evidence-based innovation. We address each of these issues in turn.

*1) Technical Barriers to Collecting and Sharing Data:* Like many areas of biological and clinical data science, creating shared data sets that are easily accessed and which are usable for data-driven research and improvement faces numerous challenges around data quality, data provenance, data scale, and data meaning. These are often competing objectives. For example, video data of minimally invasive surgical procedures is technically easy to access and to store, however it is difficult to use as it is unstructured and it has high volume. Furthermore, to make it usable, it must be combined with other external data that creates a link between what can be measured in the video, and elements of clinical value. This external data may itself be clinical data or imaging data, which is also challenging to access and normalize into a consistent format. Finally, the costs associated with acquiring and storing such a volume of the raw data may exceed the value that can be extracted, given the current state of the art.

Conversely, data that is acquired via other more intrusive technological means may be more compact and have more immediate value, but suffers from scalability issues. For example, in [96] data is acquired in the OR by manually recording which tools are used at one second intervals and automatically determining the status of the endoscope and the presence of clips. This data is then used to model the surgical workflow with promising results. Sigma Surgical (http://www.sigmasurgical.com) has created a means to record specific time-points in the OR, which can be assessed for operative efficiency. However, both of these approaches require additional instrumentation, manual intervention as well as adding time to the workflow, and therefore face challenges to adoption and scalability.

Provenance and data semantics are equally difficult problems. For example, evaluating the efficiency of a particular approach to cholesystectomy will require a standardized nomenclature for phases of the procedure. However, such a standardized nomenclature does not exist. Furthermore, defining the specifics of when a particular phase begins or ends, or other attributes of a performance, is equally ill-defined and is challenging to establish. Recently published data sets such as EndoVis[13], M2CAI16[14], Cholec80 [12] and EndoTube [118] illustrate the challenges of creating large scale consistent benchmarks.

Making progress on these problems will require establishing an interlocking set of standards, technical methods, and value points for the community. Clinical registries provide a good example of such a mechanism. In a registry, a specific area of practice agrees on data to be shared, outcome measures to be assessed, and standardized formats and quality measures for the data [119]. Identifying areas of SDS where the value proposition exists to drive the use of registries would provide a much-needed impetus to create data archives.

*2) Tools and Methodologies:* The goal of SDS is ultimately to improve the value (quality and efficiency) of surgery. As noted in [117] a key element for the field is to establish community metrics, and to define what level of reproducibility the field expects for these measures. One way to make such measures and methodologies concrete is to create standardized tools and practices associated with the data. For example the JIGSAWS data set [120], [121] include data, data labels, a published methodology with baseline results, and the code and tools that were used to generate those baselines. Similar archives such as the Middlebury Stereo Benchmark [122] provide a standardized set of data, tools, and evaluation metrics.

Standardized data collection tools, tied to metrics, are also key to SDS. For example, creating easy-to-use, workflow-consistent tools to capture data in the OR, and similar tools to harvest associated clinical data from EMRs or (more recently) patient-collected data related to outcomes will greatly accelerate progress. To this end, engaging companies to create open interfaces and standardized data exchange protocols would greatly reduce the barriers to collection and accelerate the development of tools and methodologies.

---

[13]http://endovissub-workflow.grand-challenge.org
[14]http://camma.u-strasbg.fr/m2cai2016/

TABLE III
SELECTION OF PUBLICLY ACCESSIBLE AND ANNOTATED SURGICAL DATA REPOSITORIES

| Source | Surgical platform | Tasks / Procedures | Data Type | Data set size | Reference annotations |
|---|---|---|---|---|---|
| JIGSAWS[1] | Robotic minimally invasive surgery | Suturing, knot-tying, needle passing | Kinematics, video | 39, 36, and 28, respectively, for the listed tasks | Skill, activity |
| ATLAS Dione[2] | Robotic minimally invasive surgery | Ball placement, ring peg transfer, suture pass, suture and knot tie, urethrovescial anastomosis | Video | 86 | Activity, tool, skill |
| Cholec80[3] | Laparoscopy | Cholecystectomy | Video | 80 | Activity, tool presence |
| m2cai16-workflow[3] | Laparoscopy | Cholecystectomy | Video | 41 | Activity |
| m2cai16-tool[3] | Laparoscopy | Cholecystectomy | Video | 15 | Tool |
| No name[4] | Endoscopy | GI endoscopy | Video | 10 | Region tracking |
| No name[4] | Robotic minimally invasive surgery | Partial nephrectomy | Video | 40000 pairs of images | None |
| NBI-InfFrames[5] | Endoscopy | Laryngolscopy | Video | 720 images | Informative frames |
| Nephrec9[6] | Robotic minimally invasive surgery | Partial nephrectomy | Video | 9 | Activity |
| EndoAbs[7] | Robotic minimally invasive surgery | Abdominal cavity visualization | Video | 120 pairs of images | Stereo 3D reconstruction |
| TrackVes[8] | Endoscopy | Abdominal cavity visualization | Video | 9 (3 ex-vivo, 6 in-vivo) | Soft tissue and context (e.g., safety area visible, presence of smoke) |
| RMIT[9] | Microscopic surgery | Retinal surgery | Video | 1500 images | Instrument position and size |
| Laparoscopy Instrument Sequence[9] | Laparoscopy | Unclear | Video | 1000 images | Location of each part of instrument |
| Pelvic Image Sequence[9] | Laparoscopy | Pelvic surgery | Video | 1 image sequence | Location of each part of instrument |
| Spine Image Sequence[9] | Microscopic surgery | Spine surgery | Video | 1 image sequence | Location of each part of instrument |
| TMI Dataset[10] | Laparoscopy | ex-vivo | Video, CT surface model | 35 stereo images, corresponding surface model | Stereo 3D reconstruction, disparity map |
| Crowd-Instrument[10] | Laparoscopy | Adrenalectomy, Pancreas resection | Videos | 120 images | Location of instrument |
| EndoVis-Instrument[11] | Laparoscopy + Robotic minimally invasive surgery & Colorectal surgery | ex-vivo | Video | 6+6 sequences | Location of each part of instrument, tool center point, 2D pose |
| EndoVis-RobInstrument[12] | Robotic minimally invasive surgery | different porcine procedures | Video | 8 | Location of each part of instrument |
| EndoVis-Workflow[13] | Laparoscopy | Colorectal surgery | Video, device signals | 30 | phase, tool |
| EndoVis-GIANA[14] | Endoscopy | Colonoscopy, Wireless Capsule Endoscopy | Video | 36 videos, 3312 images | polyp/angiodysplasia masks, classification |
| EndoVis-RobSeg[15] | Robotic minimally invasive surgery | Nephrectomy | Video | 8 | Location of each part of instruments, objects, anatomy |
| CATARACTS[16] | Microscopy | Cataract surgery | Video | 100 | tool |

[1] https://cirl.lcsr.jhu.edu/research/hmm/datasets/jigsaws_release/
[2] https://www.roswellpark.org/education/atlas-program/research-development/dione-dataset
[3] http://camma.u-strasbg.fr/datasets
[4] http://hamlyn.doc.ic.ac.uk/vision/
[5] https://zenodo.org/record/1162784#.WvWlTmaZNN2
[6] https://zenodo.org/record/1066831#.WvWlYmaZNN1
[7] https://zenodo.org/record/60593#.WvWlcGaZNN1
[8] https://zenodo.org/record/822053#.WvWlgWaZNN1
[9] https://sites.google.com/site/sznitr/code-and-datasets
[10] http://open-cas.org/
[11] https://endovissub-instrument.grand-challenge.org/
[12] https://endovissub2017-roboticinstrumentsegmentation.grand-challenge.org/
[13] https://endovissub2017-workflow.grand-challenge.org/
[14] https://endovissub2017-giana.grand-challenge.org/
[15] https://endovissub2018-roboticscenesegmentation.grand-challenge.org/home/
[16] https://cataracts2018.grand-challenge.org/

*3) Barriers to be Overcome:* There are numerous regulatory, technical, and sociological barriers that inhibit data set creation. Concerns about patient and surgeon privacy and (in the US) potential liabilities present impediments to large-scale data aggregation. The impact of these issues varies, however, depending on problem setting. Acquiring hand motion data from surgeons in training on benchtop phantoms presents almost no difficulty. Acquiring video data from live open surgery presents numerous challenges due to the challenges of de-identification.

A key step will be to identify areas where progress is possible, and where value can be established. A possible solution path is to create categories of data and collection settings, and then create associated standards for acquisition, anonymization, and publication of data. Having consistent and workable models will allow the community to exhibit workable models that will provide ethics boards a "template" which will make it easier to get new data collections established.

Sociological barriers also exist. Ultimately, data should be collected as a matter of best-practice in a consistent, longitudinal manner – just as with any college or professional athlete. To do so, surgeons and surgical teams will need to be active and engaged participants in SDS. However, current medical school and residency training does not include the use of data science approaches. There are no value models or incentives that drive interest. An as yet unexplored challenge and opportunity is the longitudinal collection of data which stretches from training to surgical practice, and ultimately includes patient data and patient outcomes. Federating these data sources will require a level of commitment and discipline that can only come with the complete "buy-in" of the medical profession. In summary, there are many opportunities to create shared data sets, however, the field needs to identify allies and clear short-term "wins" that will build interest and trust in the area so that hospitals, insurers, and practitioners all see the value of creating the resources to advance the profession.

*E. Towards shared (open-source) tools*
Author: D. Stoyanov

The field is inherently underpinned by the convergence of different specializations, for example natural language processing to mine patient records; medical image computing generating segmentations, models, atlases of the patient anatomy; wearables and devices that track physiological signals; and surgical instrumentation including robotics and interventional suite apparatus. This means that open-source projects from each of these technical specialties, for example medical image computing, have a role towards realizing such initiatives in SDS. This section discusses the World Café Question W5: "Shared tools and software: What do we need? What do we have?" (Tab. II-B).

*1) Existing open source initiatives:* The computing community in many related fields has established practices for providing open source implementations of algorithms alongside publications as well as large scale software projects integrating software solutions. There are many relevant projects, often interconnected by common libraries, to help joint functionality, format standardization, rendering or data transfer.

*a) Medical image computing software:* : Researchers in medical image computing and computer assisted interventions have a long history of open source code platforms. Common early tools were built to make it possible to load the rather complex DICOM format which encapsulated medical image data but can vary in implementation from different vendors of imaging scanners. With the evolution of such starting code many frameworks now embed very advanced functionality for medical image computing such as registration algorithms, segmentation or detection and classification methods. Examples of such software include 3DSlicer [123] (https://www.slicer.org), The Medical Imaging Interaction Toolkit - MITK [124] (http://mitk.org), Nifty Tools [125] (http://cmictig.cs.ucl.ac.uk/research/software/software-nifty) and the Insight Segmentation and Registration Toolkit - ITK (https://itk.org).

*b) Computer vision:* Similarly to activities for medical image computing, in computer vision open source development and code sharing is well established. Apart from multiple open source projects that provide specific algorithm implementations, the largest frameworks in widespread use are the Open Computer Vision Library - OpenCV (http://opencv.org) and VLFeat (http://www.vlfeat.org). Large projects such as the VXL (the Vision-something-Libraries) (http://vxl.sourceforge.net) that were once popular seem to have dropped from maintenance. Notably vision open source libraries typically feature implementations of video opening, seeking and encoder/decoder handling code such as FFmpeg (https://ffmpeg.org) which historically suffered from platform dependence but have evolved into cross-platform tools with the growing support from hardware manufacturers to handle different OS platforms for both cameras and capture devices. Notably for computer vision research, MATLAB toolboxes are also extremely popular tools now with the platforms improved computational capabilities and ease of prototyping of complex algorithms.

*c) Machine learning:* The growing influence on machine learning approaches, recently deep learning in particular, is profound in both the medical image computing and the computer vision fields. Software instruments from machine learning have emerged in different platforms that enable handling of large volumes of data for training. Popular examples at the moment include Caffe (http://caffe.berkeleyvision.org), TensorFlow (https://www.tensorflow.org), Torch (http://torch.ch) and Theano (http://deeplearning.net/software/theano). The importance and power of these systems cannot be understated and they are currently used to design the most effective algorithms for model free processing in most classic problems encountered in computer vision and in medical imaging. With the growing capability and labeling of medical image data and other OR generated data these platforms will continue to flourish in SDS.

*d) Robotics:* The Robot Operating System - ROS (http://www.ros.org) has matured into the most widely used platform for developer level and lab robotic systems. The main reason for this has been the community development of connections to robots from the main manufacturers of industrial platforms as well as educational devices. It is widely used in research and teaching and typically also for industrial development prior to product production. ROS integrates imaging platforms like OpenCV and other libraries or tools like for example Gazebo for simulation. The counter to all the advantages to ROS is that it is a heavyweight system that is not cross platform at present and cannot be used in truly real-time mode. For surgical robots this is currently a barrier but not relevant to research systems.

It should be noted that the exemplar open source initiative which are mentioned above typically have cross dependences of various kinds or are included in common bundles. They may share underpinning numerical libraries for linear algebra or optimization, likewise system functionality may be provided by frameworks like Boost (http://www.boost.org) and GPU support by libraries developed for specific hardware.

*2) Existing platforms:* Various machine learning, cloud and cluster computing solutions are now available from large providers such as Microsoft, Google, Amazon and others. Additionally toolboxes as part of development environments such as MATLAB are becoming available albeit with license fees in addition to research led open source code repositories.

*3) Specialist tools that are needed:* Video is typically used as a means for capturing activity information at different granularities ranging from cameras observing the whole interventional room or suite to cameras inserted into the body endoscopically or observing specific sites through a microscope. The power of video information is that it embeds multiple SDS cues such as the motion and communication of the theatre staff or the dexterity and surgical skill competence of instrument manipulation. Annotation, especially for video data, allows consistent and easy meta-data labelling of videos. Signal synchronization is required to be able to do multivariate analysis. Time series analysis.

*4) Challenges:* We summarize the main challenges for open source initiatives for SDS, some of which are shared with those that open data initiatives need to overcome.

*a) Funding and maintenance:* Government led organizations are predominantly focused on funding scientific innovation. The importance of software design and engineering, maintenance, management and development are not currently recognized as priority areas. This makes it very difficult for research teams and organizations to build and sustain their core technology and code base which is predominantly software based. Frameworks tend to quickly demise when insufficient support and core effort can be put into maintaining builds and documentation, as well as support to help maintain momentum in the community.

*b) Controlling size and scalability:* Retaining knowledge bases, documentation and training are critical for large open source frameworks. Existing initiatives, such as Slicer and MITK, have made great efforts to host hackathons and joint workshops to share knowledge and provide training. This needs to be maintained and becomes additionally important and difficult as projects grow in size. The natural result of greater capabilities and functionality is complexity, which elongates learning curves for new users and can be a barrier to entry if not properly mitigated.

*c) Regulatory and standardization issues:* For systems developed to empower SDS to be used, they need to be able to provide value before, during or after surgery or interventions. Validating and verifying such capabilities requires trials in real treatment practices and these must be regulated and approved by relevant bodies in terms of ethics to ensure safety and proper process. System documentation and reliability is critical to pass through such approval procedures but can be obtained for research purposes without proof of code stability. On the other hand, once a system or capability needs to be transitioned into widespread use or even before to late phase trials, certification becomes an important consideration. Certification varies across different geopolitical zones but typically requires strict processes to be followed during code development and may require separation of certain open source code blocks that do not comply with the requirements of the body and process.

*F. Towards standardized validation*
Authors: P. Jannin and S. Vedula

This section discusses the World Café Question W6: "Validation: How can we ensure validity of findings in surgical data analysis?" (Tab. II-B).

SDS requires strong validation and evaluation as much as, if not more, as it is needed in computed aided surgery. It is required for many reasons. SDS is an emerging field. As so, it needs to demonstrate its validity and added value to convince the actors of health care including stakeholders and patients. As part of the new emerging digital world, SDS aims to impact decision-making and health care. It should demonstrate its ethical dimension and not be considered as an opaque system that suggests decisions to physicians, according to opaque criterion and for the benefit of unknown actors. In order to do so, high quality of results, transparency of processes through common and available methods and data are required. Validation or validity, as defined as the demonstration that a system does what it has been designed to do, as well as evaluation, as defined as the demonstration of the short, mid and long term added values of the system, are both needed. What additionally makes validation and evaluation (VE) complex in such an area is that it should concern SDS methods, SDS models built from methods, systems built with SDS models and methods, and applications built from such systems. Several issues related to VE were emphasized during the SDS workshop, such as access to validation data, standardization of VE, application relevant objectives, and multidisciplinarity of VE.

Access to validation data was the topic cited the most by participants. There are several issues regarding validation data. First, the validation data should be reliable, precise, and accurate. In one word, validation data, including the associated reference, should be valid, in the sense of validity. Such data should reflect the huge variability existing in surgery, as mentioned above, including rare cases. Both local/specialized and global/generic references are required. Significance of the results should be guaranteed by a large number of samples. Not only is high quality data required, but also a lot of data. In surgery, being able to generate such clinical data for validation with ethical approval is difficult and costly as was also discussed in the workshop *Comment* [3].

In order to ensure proper validation and evaluation, it has already been outlined that standardization is of crucial importance: standard data or benchmarks on which validation is performed, standard metrics to estimate quantitative validation and evaluation, and standard VE processes and methodologies[126]–[128]. For all, standardization helps comparison of results across studies and may reduce costs and time from bench to bedside. To start again with data, low biased and highly realistic data are needed, requiring access to different kind of data from simulated to clinical ones. Ethical approval is needed. Standardized data should be freely available and well documented from opendata or challenge schemes. Similarly, standard metrics and methodologies should be freely available and well documented. Both technical and clinical based metrics are required. Globally all levels of validation and assessment need to be covered with available and standard metrics. A special emphasis on clinical outcomes was requested by the participants. The relevance of the VE studies with regards to clinical applications has been outlined as crucial in SDS. Application dependent expectations, defined as clinically meaningful research questions or objectives, along with expected values are expected. Additionally, guidelines have to be built explaining methodologies to be followed for ensuring quality and relevant results in VE[127]. For data, reference and metrics, quantification of uncertainties should also be made available. Due to diversity and complexity of validation and evaluation, multidisciplinary actors should be involved including researchers, statisticians, human and social scientists, clinicians along with the corresponding clinical societies, and stakeholders from industry and public bodies.

VE in SDS is mandatory, but complex and costly. This is a continuous process that is expected along the product lifecycle. It is not obvious who can cover costs of data production and standardization, as well as their ownership. Open initiatives are preferred to ensure large dissemination and avoid conflicts of interest. Some preliminary initiatives exist already, for instance in the area of biomedical challenges. They have to be further discussed, extended and finally adopted by all of the involved actors.

### G. Incentives for advancing the field
*Author: C. Pugh*

The current section is based on World Café Question W7: "How can we convince clinicians and other stakeholders to invest in the SDS?" (Tab. II-B).

A top priority in advancing the field of SDS is improving the quality of care provided to patients. High quality care can be achieved by strategically positioning hospital systems to become more efficient, safe, reliable and cost effective. Achieving this goal will provide benefits to numerous stakeholders including but not limited to patients, healthcare providers, hospital administrators and investors. So how do we convince clinicians and other stakeholders to invest in SDS? The key area that must be addressed is defining and creating value for the stakeholders.

*1) Incentives for Patients:* One of the most important drivers for patients seeking healthcare is perception. While SDS could greatly facilitate high quality, safe and efficient care, it will not be successful if patients do not perceive that they will derive benefit. For example, advancements in robotic surgery for urologic procedures have well-documented, proven benefits. However, while these benefits remain unproven for a variety of other surgical procedures, most patients still believe that robotic surgery is superior. To address patient perception, SDS will need to appear superior to current methods of measuring and communicating patient safety. Current measurement areas to target may include physician board certification, public healthcare provider ratings and hospital quality ratings. In addition, an untapped area includes providing patients with more and better information about themselves. Patient access to medical records is a hot topic [129]–[131]. While the nuances of providing full access have not been sorted out, what remains clear is that patients want more information about their health, treatment and outcomes and SDS has potential to enable a clear and efficient way of accomplishing this goal by sorting out what patients want to know and presenting a dashboard or menu of personalized viewing options.

*2) Incentives for Clinicians:* There are a number of drivers for clinicians. However, trust in the data that may be generated from SDS is a very high priority. Is the data reliable? How will it be used? These are a few of the questions that must be addressed. Other potential benefits for clinicians include workflow incentives such as efficient patient management, decision support and personalized feedback on clinical skills which may enact a competitive advantage. As clinicians are increasingly burdened with laborious chart documentation, it appears that there are numerous opportunities to use SDS concepts and approaches to facilitate the chart documentation process and help to ensure that the most pertinent diagnostic and treatment information are not lost or buried in an unsearchable format or one that is difficult to access, process or understand.

*3) Incentives for Hospital Systems:* Creating value for hospital systems means addressing the business of healthcare and financial viability. SDS is well positioned to facilitate the lean approach. Lean hospitals use data and strategic planning to get the highest return on their investments and larger profit margins. Hospital leadership with a background and training in SDS would be greatly beneficial in moving the agenda forward by being in a position to drive the learning, implementation and evaluation of benefits.

*4) Incentives for the Surgical Data Scientist:* Research funding and recognition as a valid and important career path are high level drivers for the surgical data scientist. Additional drivers include collaboration, data sharing and advancing the science.

*5) Incentives for Research Based Funding Agencies:* Investment in a new scientific area that has great potential to

advance current thinking is the major incentive for research based funding agencies.

*6) Summary:* A common theme in defining the incentives for stakeholders to invest in SDS is value. Each stakeholder group has a different set of values. Funding agencies and data scientists must work together not only to advance the science at the most basic and fundamental levels but also to translate the science from a methodological, theoretical and algorithm based endeavor into clear and tangible benefits for patients, healthcare providers and hospital systems. Further thoughts on the incentives and views of the different stakeholders are presented in the published *Comment* [3] on the workshop.

### H. Dissemination strategy
Author: A. Park

One World Café group assignment was to consider what the "product" (i.e., the collaborative efforts of Physicians and Data Scientists) coming out of this might look like and how it would most effectively be disseminated/distributed (W8: "Dissemination : How can we translate research results into clinical practice?" (Tab. II-B)).

There is little if any peer reviewed literature to inform discussion of our assigned topics. Several high level observations that could be a challenge to SDS product dissemination were made and widely agreed upon. Chief among them a real world (if not cynical) view that the medical technology business – as presumably with other business fields – is replete with examples of sales and marketing forces trumping substantive products derived from solid research and development efforts that have insufficient marketing "critical mass"!

Ultimately the successful distribution of a product may depend far more upon the size of a company and its sales and marketing infrastructure than the brilliance and effectiveness of the idea/product. Thus far products of nascent SDS collaborative efforts, as clever, durable and effective as they may be, face significant barriers to dissemination unless acquired by a "major player".

A related observation of the group was that the focus of the field of SDS would likely shift significantly, conceivably to the point that those who "birthed" it would lose control of any higher order agendas, were software behemoths such as Google, Apple or Microsoft to focus their attention and resources in this direction. It was acknowledged by group members that industry "players" in the surgical device communication integration space are incentivized to develop proprietary systems and products rather than open source software and systems of interchangeable devices that surgeons and hospital administrators would obviously prefer. A key point to consider and discuss with all potential stakeholders is how the move internationally to "fee for value" from "fee for service" health care, will provide opportunities to align incentives among all parties and thus spur investment. There are few models of effective dissemination of SDS type products or surgical devices in the absence of meaningful licensing or distributing agreements with major industry players.

In the most elemental terms, the group posited that any viable product of SDS efforts must conform to the following design specifications, it must:

- Improve patient safety and quality of care, i.e., outcomes
- Improve patient satisfaction, i.e., patient experience
- Reduce cost

Several discreet product ideas to be evolved from SDS work were proposed by the group:

1) Development of a surgical equipment/device/resource and patient tracking system to manage surgical assets more efficiently (equipment, OR block time, etc.) and enhance patient flow (and satisfaction) through the cycle of surgical care. Such a system would need of course to be "smart" and capable of reliable, predictive modeling (of OR utilization, asset needs, etc.) and decision support.
2) System to objectively monitor surgeon maintenance (and decay) of clinical judgment and technical skills by evolving high stakes (validated) assessment of the surgeon and then offering personally "prescribed" or tailored skills modules as remediation per surgeon.
3) Cost savings programs that would provide decision support specifically to identify savings opportunities with OR cases that are high volume, high cost with high cost and outcome variability across a number of surgeons performing the same procedure.
4) Develop a system to track patients within an entire medical system (not just OR) and provide real time info to the patient and their care givers with regard to current location, intended destination, constantly updating and presenting relevant PHI data from EMR, PACS, etc. in real time.

Clearly the theme among proposed products leans strongly to decision support with varying degrees of scope/reach and sophistication. Such product ideas will emerge as physicians and data scientists work more closely together on surfacing unmet needs and creating their solutions. Further thoughts of the SDS board with respect to dissemination and impact are summarized in [3].

## VII. CONCLUSION

This document is a summary of the first international workshop on SDS, held in Heidelberg in June 2016. Based on the content presented herein, a *Nature Biomedical Engineering* article [3] was produced that presents a definition of SDS as well as a concise summary of the key challenges associated with the field. The main conclusions of the workshop can be summarized as follows [3]:

- "SDS will pave the way from artisanal to data-driven interventional healthcare with concomitant improvements in quality and efficiency of care.
- A key element will be to institutionalize a culture of continuous measurement, assessment and improvement using evidence from data as a core component.
- An actionable path would be that societies support and nurture efforts in this direction through best practice, comprehensive data registries, and active engagement and oversight.
- SDS should be established as a new element of both the education and career pathway for hospitals that teach and train future interventionalists."

*A. Acknowledgements*

The authors thank the Transregional Collaborative Research Center (SFB/TRR) 125: Cognition-Guided Surgery, funded by the German Research Foundation (DFG), for sponsoring the workshop that served as basis for the manuscript (www.surgical-data-science.org/workshop2016). We also thank Estuardo Calderón Scheel (Project Solutions GmbH) for moderating the SDS workshop and N. L. Rodas, M. A. Cypko and all other workshop participants for their valuable input during the workshop. Thanks also to Xinyi (Cindy) Chen for compiling information presented in Table VI-D. Finally, we acknowledge the support of the European Research Council (ERC-2015-StG-37960), French Investissements d'Avenir program (ANR-11-LABX-0004, ANR-10-IDEX-0002-02), the US National Institutes of Health (NIH R01EB01152407S1, NIH/NIBIB P41 EB015902, NIH/NCI U24CA180918, NIH/NIBIB P41 EB015898, NIH/NIBIB R01EB014955,NIH/NIDCR R01-DE025265), the US Department of Defense (DOD-W81XWH-13-1-0080), the Royal Society (UF140290), the French Investissements d'Avenir program (ANR-11-LABX-0004, ANR-10-IDEX-0002-02)," and the Link Foundation Fellowship in Advanced Simulation and Training.